\documentclass[aps,prb,reprint,groupedaddress, showpacs,citeautoscript]{revtex4-1}

\usepackage{amsmath,amssymb,graphicx}
\usepackage{dsfont}

\usepackage{extarrows}
\usepackage{todonotes}

\newcommand{\op}[1]{\hat{#1}}
\newcommand{\up}{\uparrow}
\newcommand{\down}{\downarrow}

\begin{document}

\title{Controlling observables in time-dependent quantum transport}

\author{K.J. Pototzky and E.K.U. Gross} 

\affiliation{Max Planck Institute of Microstructure Physics, 
             06120 Halle (Saale), Germany}

\date{\today}

\begin{abstract}
    The theory of time-dependent quantum transport
    addresses the question: How do electrons flow through a junction
    under the influence of an external perturbation as time goes by? 
    In this paper, we invert this question
    and search 
    for a time-dependent bias such that the system
    behaves in a desired way. 
    This can, for example, be an observable
    that is forced to follow a 
    certain pattern 
    or the minimization of an objective 
    function which depends on the observables.
    Our system of choice
    consists of quantum dots coupled to normal 
    or superconducting leads. 
    We present results for junctions with normal
    leads where the current, the density or 
    a molecular vibration
    is optimized
    to follow a given target pattern. 
    For junctions with two superconducting leads, where
    the Josephson effect triggers the
    current to oscillate,
    we show how to suppress the Josephson oscillations
    by  suitably tailoring the bias.
    In a second example involving superconductivity, we consider a Y shaped junction 
    with two quantum dots coupled to one superconducting and two normal leads.
    This device is used as a Cooper pair splitter to
    create entangled electrons on the two quantum dots.
    We maximize the splitting efficiency
    with the help of an optimized bias.
\end{abstract}


\maketitle

\section{Introduction}
    \label{sec:Introduction}
        
    Molecular quantum transport
    is a fast growing research field.
    The ultimate goal is
    to produce electronic devices
    using single molecules
    as their building blocks \cite{Aviram1974, Cuniberti2005, DiVentra2008, Cuevas2010}.
    The prospective improvements
    regarding operational speed as
    well as storage capacity 
    are expected to be enormous if the
    miniaturization
    of transistors can be taken to the scale
    of single molecules.
    
    In the past, the main objective 
    was to measure and/or calculate
    the current-voltage characteristics
    of the molecular junction.
    On the theory side, calculations were 
    usually done
    within the Landauer-B\"uttiker approach.
    In recent years, interest has shifted 
    more and more towards time-resolved 
    studies. Such studies allow one to 
    address questions like: How long does it
    take until the steady state is reached?
    Can we shorten or lengthen this time span? 
    Does a steady state always exist, and if so,
    is it unique?
    To answer this kind of questions by calculations,
    explicitly time-dependent
    approaches are 
    necessary, such as
    time-dependent density functional theory 
    \cite{RungeGross1984, Baer2004, Burke2005, Kurth2005, YuenZhou2009, DiVentra2007, Appel2009, Appel2011},
    the Kadanoff-Baym equations \cite{Stefanucci2013,Myoehaenen2009, Myoehaenen2010, Knap2011},
    multi-configuration time-dependent
    Hartree-Fock \cite{Zanghellini2004, Wang2009, Wang2013, Albrecht2012},
    Quantum Monte-Carlo \cite{Muehlbacher2008},
    time-dependent tight binding \cite{Wang2011b,Zhang2013,Oppenlaender2013},
    or the hierarchy equation of motion approach \cite{Jin2008, Zheng2008, Zheng2008b}.
    
    In all those approaches the reaction of the molecular
    junction 
    to a given external perturbation, i.e. a bias or a gate voltage
    is calculated.
    In this article, 
    we want to take a step beyond this point and
    control the current or other observables
    of the junction. This means
    we have to address the inverse question:
    Which perturbation leads to a desired reaction
    of the system? To answer this question,
    optimal control theory
    provides a suitable framework.
    This research field
    was pioneered by the work of Pontryagin \cite{Pontryagin1962} 
    and Bellman \cite{Bellman1957}
    who paved the way for numerous
    applications. Initially, 
    optimal control theory was mainly
    used to solve problems of classical 
    mechanics.
    Later, it found applications
    in many other research fields
    including quantum mechanics. \cite{Peirce1988,Shi1988,Kosloff1989}

    A particularly interesting field goes under 
    the heading of ``femto-chemistry''
    where chemical reactions are
    influenced with femto-second 
    laser pulses such that
    a specific reaction gets
    suppressed or enhanced. \cite{Assion1998,Hartke1989,Elghobashi2003,Elghobashi2004}
    A successful experimental 
    application is the 
    selective bond dissociation
    of molecules. \cite{Levis2001}
    Other applications of optimal control
    theory in the quantum world include
    the control of the electron flow 
    in a quantum ring \cite{Raesaenen2007},
    the accelerated cooling 
    of molecular vibrations \cite{Reich2013},    
    the control of the entanglement 
    of electrons in quantum wells \cite{Raesaenen2012},
    the optimization of quantum revival \cite{Raesaenen2013b},
    the control of ionization \cite{Castro2009,Raesaenen2012b}
    or the selection of
    transitions between molecular states \cite{Jakubetz1990}.

    Kleinekath\"ofer and coworkers combined
    optimal control theory with the master
    equation approach for quantum
    transport and demonstrated the 
    control of various observables in junctions 
    with normal leads \cite{Li2007, Amin2009, Li2010}.    
    We take a different approach to the same problem
    by propagating wave functions.
    For the time propagation, we employ an algorithm    
    proposed by Stefanucci \textit{et al.} \cite{Stefanucci2010}.
    This allows us to treat not only normal 
    (N) but also superconducting (S) leads.
 
    The paper is organized as follows: 
    In section \ref{sec:Model}, we
    explain the model 
    Hamiltonian 
    that we employ to describe the molecular junctions. 
    In section \ref{sec:optimization_problem}, we 
    formulate the optimization problem for 
    tailoring the bias such that a chosen observable
    follows s predefined pattern as best as possible. 
    Various results are presented in section \ref{sec:results}.
    Finally, 
    in section \ref{sec:optimize-splitting},
    we focus on a specific example,
    a Y shaped junction
    consisting of two quantum dots coupled to
    one superconducting and two normal leads.
    This device is used as a Cooper-pair splitter, for which we 
    maximize the splitting efficiency.
    In the final section \ref{sec:conclusion}, we draw our conclusions.

\section{Model}
    \label{sec:Model}
        Our model system consists of 
        a quantum dot (QD) connected to two semi-infinite, non-interacting
        one dimensional leads (L and R), which are described by a tight binding Hamiltonian. 
        Later, in section \ref{sec:optimize-splitting}, we will
        add a third lead (labeled S) and a second quantum dot.
        The corresponding changes in the Hamiltonian 
        will then be stated in that section
        but the overall approach and 
        the structure of the equations stays the same.

        The Hamiltonian for the junction with two leads
        and a single quantum dot reads
        \begin{align}
            \label{eqn:sec2:full-Hamiltonian}\op{H}(t)            &= \op{H}_{\textnormal{QD}} + \sum_{\alpha \in \{\textnormal{L},\textnormal{R}\}} \op{H}_{\alpha} + \sum_{\alpha \in \{\textnormal{L},\textnormal{R}\}}\op{H}_{T,\alpha}(t)
        \end{align}
        with
        \begin{align}
            \op{H}_{\textnormal{QD}}     &= \epsilon_{\textnormal{QD}} \sum_{\sigma\in\{\up, \down \}} \op{d}_{\sigma}^\dagger\op{d}_{\sigma},\\
            \op{H}_{\alpha}              &= \sum_{k=0}^\infty \sum_{\sigma\in\{\up,\down\}}\left(t_\alpha \op{c}_{\alpha k\sigma}^\dagger \op{c}_{\alpha(k+1)\sigma} + H.c. \right),\\
                                         &  \nonumber\qquad + \sum_{k=0}^\infty \left(\Delta_\alpha e^{i\chi_\alpha} \op{c}_{\alpha k\up}^\dagger \op{c}_{\alpha k\down}^\dagger + H.c. \right),\\
            \op{H}_{T,\alpha}(t)         &= \sum_{\sigma \in \{\up,\down\}}\left( t_{\alpha, \textnormal{QD}} e^{i\gamma_{\alpha, \textnormal{QD}}(t)}\op{c}_{\alpha0\sigma}^\dagger \op{d}_\sigma + H.c. \right).
            \label{eqn:sec2:last-eqn-full-Hamiltonian}
        \end{align}
        Here $\gamma_{\alpha, \textnormal{QD}}(t) = \int_0^t\,dt' U_\alpha(t')$
        are the Peierls' phases with the bias $U_\alpha(t), \alpha \in \{\textnormal{L},\textnormal{R}\}$.
        The operator $\op{c}_{\alpha k\sigma}^\dagger$ ($\op{c}_{\alpha k\sigma}$) creates (annihilates) 
        an electron at site $k \in \mathbb{N}$ in the lead $\alpha \in \{\textnormal{L},\textnormal{R}\}$ with spin $\sigma \in \{\up,\down\}$.
        The operator $\op{d}_\sigma^\dagger$ ($\op{d}_\sigma$) 
        represents the creation (annihilation) of an electron on the quantum dot.

        The observables of prime interest,
        the
        density $n_{\textnormal{QD}}(t)$ and the current $I_{\alpha, \textnormal{QD}}(t)$,
        are given by
        \begin{align}
            n_{\textnormal{QD}}(t)         &= \sum_{\sigma \in \{\up, \down\}}\langle \op{d}_{\sigma}^\dagger (t) \op{d}_{\sigma}(t) \rangle,\\
            I_{\alpha, \textnormal{QD}}(t) &= 2\Im \sum_{\sigma \in \{\up,\down\}} \left(t_{\alpha, \textnormal{QD}}e^{i\gamma_{\alpha, \textnormal{QD}}(t)} \langle \op{c}_{\alpha 0\sigma }^\dagger(t) \op{d}_\sigma(t) \rangle \right).
        \end{align}
        
        All parameters in equations (\ref{eqn:sec2:full-Hamiltonian}) - (\ref{eqn:sec2:last-eqn-full-Hamiltonian}) are real and positive.
        We always work at temperature $T=0$ and in the wide band limit $t_{\alpha, \textnormal{QD}} \ll t_\alpha $, where
        the coupling to the leads is given by
        $\Gamma=\Gamma_\textnormal{L}+\Gamma_\textnormal{R}, \Gamma_\alpha = \frac{2t_{\alpha, \textnormal{QD}}^2}{t_\alpha}, \alpha \in \{\textnormal{L},\textnormal{R}\}$.
        In this limit, the results only depend on the couplings $\Gamma_\alpha$ but
        not on the hopping elements individually.
        The superconducting pairing potentials $\Delta_\alpha$ can be written as $\Delta_\alpha = \xi_\alpha \tilde{\Delta}$,
        which allows a dimensionless representation of the problem by measuring
        times in units of $\tilde{\Delta}^{-1}$ and 
        energies as well as currents in units of $\tilde{\Delta}$. 
        In the case of normal leads, we set $\xi_\alpha=0$,
        and $\xi_\alpha=1$ otherwise.
        The presence of superconductivity
        requires the use of the 
        time-dependent Bogoliubov-de Gennes equation
        \begin{align}
            \label{eqn:BdG-equation}
            i\frac{\,d}{\,dt}
            \left(
                \begin{matrix}
                    u_q(k,t)\\
                    v_q(k,t)
                \end{matrix}
            \right)
                &=
            \sum_l
            \mathbf{H}_{kl}(t)
            \left(
                \begin{matrix}
                    u_q(l,t)\\
                    v_q(l,t)
                \end{matrix}
            \right),\\
            \mathbf{H}_{kl}(t) &=
            \left(
                \begin{matrix}
                    \mathbf{h}_{kl}(t)           & \mathbf{\Delta}_{kl} \\
                    \mathbf{\Delta}_{kl}^\dagger & -\mathbf{h}_{kl}^\dagger(t)
                \end{matrix}
            \right).
        \end{align}
        
        The single-particle wave functions
        \begin{equation}
            \psi_q(k,t) = [u_q(k,t), v_q(k,t)]^t
        \end{equation}
        represent the time-dependent particle- and hole-amplitudes
        at site $k$.
        The algorithm for the time propagation of the single particle
        wave functions $\psi_q(k,t)$ as well as the 
        initial state calculation is explained in the
        work of Stefanucci \textit{et al.} \cite{Stefanucci2010},
        which extends the method of Kurth \textit{et al.} \cite{Kurth2005}
        to superconducting leads.

    \section{Optimization problem}
        \label{sec:optimization_problem}
        We start at $t=0$ in the ground state
        of the junction with $U_\alpha(t\le 0)=0$. The goal is  
        to tailor the bias $U_\alpha(t)$ such
        that the observable of choice $O(t)$ follows
        a predefined target pattern as best as possible.
        The corresponding optimization problem reads
        \begin{align}
            \label{eqn:OCP}
            \min_{U_\textnormal{L}(t),U_\textnormal{R}(t) }  \| O[\Psi](t) - O^{(\textnormal{target})}(t) \|_{2, [0,T]}^2 \quad\quad \\
            \begin{array}{rrl}
            \text{s.t.} &  i\partial_t \psi_q(t) &= \mathbf{H} [U_\textnormal{L}, U_\textnormal{R}](t)\psi_q(t), \ t \in [0, T],\nonumber\\
                        & \quad \ \psi_q(0)      &= \psi_q^0.
            \end{array}
        \end{align}
        Here, $\|\cdot \|_{2, [0,T]}$ denotes the
        $L^2$-norm on the time interval $[0,T]$, i.e. 
        the objective function is the following integral:
        \begin{equation}
            \int_0^T \,dt | O[\Psi](t) - O^{(\textnormal{target})}(t) |^2.
        \end{equation}
        The integral is well-defined since
        $T$ and the integrand 
        are finite in all examples studied in this work.
        
        Most common is a variational approach to this problem, like
        the Rabitz approach \cite{Zhu1998}
        or Krotov's method \cite{Sklarz2002, Palao2003}.
        Such an approach incorporates the constraints into the objective function
        using Lagrange multipliers 
        and searches for the roots of the 
        variation of the new objective function.
        An alternative approach,
        which we shall adopt in this article,
        is the direct 
        minimization of the objective function
        using derivative-free 
        minimization algorithms.
        This strategy was successfully used
        in several works \cite{Castro2009, Krieger2011, Hellgren2013, Raesaenen2013}.
        In this way, we avoid various difficulties
        arising from the time propagation algorithm.
        This approach can be viewed as 
        the computational analogue to the
        closed-loop learning algorithms
        employed in experimental optimization \cite{Judson1992}.

\begin{figure}[htb]
            \begin{center}
                \includegraphics{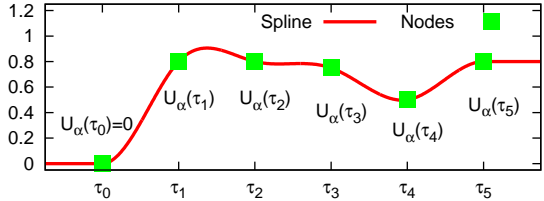}
                \caption{Cubic spline interpolation using six nodes $\tau_k$.
                         The optimization algorithm changes the values $U_\alpha(\tau_k), k \in \{1,\ldots, 5\}$.
                         The value $U_\alpha(\tau_0)$ is fixed to zero. The
                         derivatives at both ends are set to zero.
                         The spline does not necessarily take the maximum or minimum value
                         at one of the nodes. In this example, the maximum lies between $\tau_1$ and $\tau_2$.}
                \label{fig:Spline}
            \end{center}
        \end{figure}

        The basic idea of our numerical approach
        is to approximate $U_\alpha(t)$ by cubic 
        splines with $N+1$ equidistant nodes 
        at $\tau_k = \frac{k}{N}T, k\in \{0,\ldots, N\}$.
        We choose $\frac{\,d}{\,dt}U_\alpha(\tau_0) = \frac{\,d}{\,dt}U_\alpha(\tau_N)=0$
        as the boundary conditions for the splines.
        The dependence of the problem (\ref{eqn:OCP}) on the bias $U_\alpha(t)$ is replaced by
        \begin{equation}
            U_\alpha(t) \to \left[U_\alpha(\tau_0), \ldots, U_\alpha (\tau_{N})\right] \equiv \vec{u}_\alpha.
        \end{equation}
        In this way, the spline-interpolated bias 
        $U_\alpha(\vec{u}_\alpha, t)$ becomes a function of $\vec{u}_\alpha$.   
        This then yields a normal non-linear optimization problem 
        with the unknown variables $U_\alpha(\tau_k)$. We further impose
        the condition $U_\alpha(\tau_0)=0$ since the 
        bias has to be continuous and we assume $U_{\alpha}(t<0)=0$. 
        Figure \ref{fig:Spline} demonstrates this approach. 

        Additionally, we add the constraint $U_{\textnormal{L}}(t)=-U_{\textnormal{R}}(t)$ unless
        otherwise stated, since it reduces the dimensionality 
        of the optimization problem in the numerical implementation
        by a factor of two. 
        This implies 
        the constraint $\vec{u}_\textnormal{L}=-\vec{u}_\textnormal{R}$.
        The resulting non-linear optimization problem is
        \begin{align}
            \label{eqn:OCP-without-constraints}
            \min_{\vec{u}_{\textnormal{L}}, \vec{u}_{\textnormal{R}} \in \mathbb{R}^{N+1}}   \| O[\Psi](t) - O^{(\textnormal{target})}(t) \|_{2, [0,T]}^2 \quad\quad \qquad \\
             \nonumber
            \begin{array}{rcl}
                \textnormal{s.t.} \quad i\partial_t \psi_q(t)  &=& \mathbf{H} (\vec{u}_\textnormal{L}, \vec{u}_\textnormal{R}, t)\psi_q(t), \ t \in [0, T],\\
                \psi_q(0)                                      &=& \psi_q^0,\\     
                \vec{u}_\textnormal{L}                         &=& -\vec{u}_\textnormal{R}, \\
                U_\alpha(\vec{u}_\alpha, \tau_0)               &=& 0,\quad  \alpha \in \{\textnormal{L}, \textnormal{R}\}.
            \end{array}
        \end{align}

        The single particle wave functions $\psi_q(t)$ in the problem (\ref{eqn:OCP-without-constraints}) are only auxiliary variables. 
        Hence, the time-dependent Bogoliubov-de Gennes equation
        can be removed from the constraint equations for the numerical implementation.
        The objective function is then written as 
        $ \| O[\psi_q^0, \mathbf{H} (\vec{u}_{\textnormal{L}}, \vec{u}_{\textnormal{R}}, t)]( t) - O^{(\textnormal{target})}(t) \|_{2,[0,T]}^2$,
        whose evaluation requires us to solve the time-dependent 
        Bogoliubov-de Gennes equation
        in order to
        calculate the observable $O(t)$.
        
        The problem (\ref{eqn:OCP-without-constraints}) can be solved using standard 
        derivative-free
        algorithms for non-linear optimization problems.
        We use the algorithms BOBYQA \cite{BOBYQA} or COBYLA \cite{COBYLA1,COBYLA2}
        provided by the library NLopt \cite{NLopt_paper}. The former 
        one does not support non-linear constraints, but 
        converges faster compared to other tested methods. The latter algorithm
        will be used for the calculations with non-linear constraints.
        
        We point out
        that the quality of the results depends on the number of nodes $\tau_k$
        for the splines. A larger number $N$ is typically favorable for better results,
        i.e. yields a better match of the observable $O[\Psi](t)$ with its target pattern
        $O^{(\textnormal{target})}(t)$.
        But, the computational cost increases with $N$.
        Besides,
        it is not guaranteed that the obtained minimum is the global minimum
        since the used algorithms are local optimization algorithms. Thus, the results 
        may depend on the initial choice for $U_\alpha(\tau_k)$.
    \section{Results}
        \label{sec:results}
        \subsection{Current and density of a NQDN junction}
            As a first example, we show the optimization of 
            the current $I_{\textnormal{L},\textnormal{QD}}(t)$ from the left lead onto the quantum dot.
            This is done for two different numbers of spline nodes $N$.
            The case $N=4$ shows strong deviations while $N=20$ already
            yields an excellent agreement
            of the calculated current $I_{\textnormal{L},\textnormal{QD}}(t)$ with its target pattern.

            \begin{figure}[htb]
                \begin{center}
                    \includegraphics{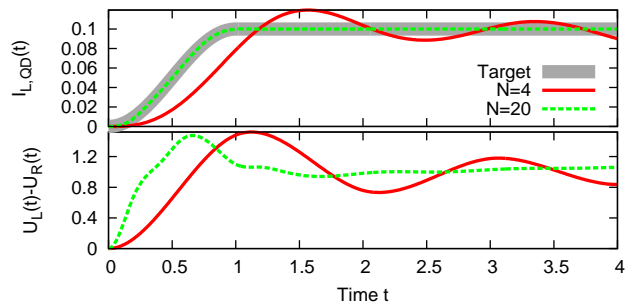}
                    \caption{NQDN junction with an optimized current for two 
                                different numbers of spline nodes $N$.  
                                The parameters are: $\Gamma_\alpha=0.2, \epsilon_{\textnormal{QD}}=0.5, \xi_\alpha=0$.}
                    \label{fig:NQDN-control-current}
                \end{center}
            \end{figure}
\begin{figure}[htb]
                \begin{center}
                    \includegraphics{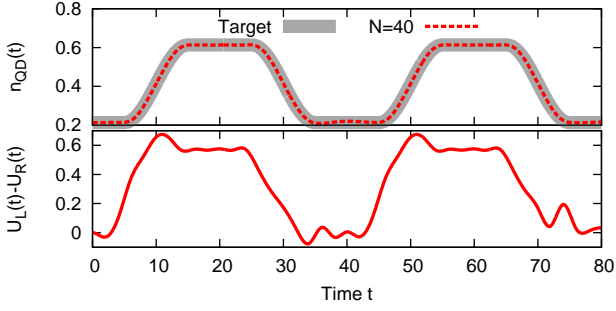}
                    \caption{NQDN junction with an optimized density. 
                            The parameters are: $\Gamma_\alpha=0.2, \epsilon_{\textnormal{QD}}=0.5, \xi_\alpha=0$.}
                    \label{fig:NQDN-control-density}
                \end{center}
            \end{figure}

            The optimization of the density $n_{\textnormal{QD}}(t)$ is very similar
            to the optimization of a current, one simply exchanges the 
            observable in the objective function. An example is shown in Fig \ref{fig:NQDN-control-density}.
            The density follows perfectly the target pattern.
    
        \subsection{Controlling classical vibrations}
            In this paragraph, we extend the model to incorporate
            a vibrational degree of freedom in the central
            region. In the past, most theoretical
            work focused 
            on the electronic
            system and neglected the nuclear motion.
            In experiments, the nuclei are, of course, not fixed to 
            a position and their
            motion can have a
            significant influence on the
            measured properties, for example on
            the current-voltage characteristics \cite{Galperin2007,Koch2005,Ryndyk2006,Haertle2009}.

            The goal of this section is to control
            the nuclear motion using 
            the bias as before. Although the bias 
            couples only to the electronic
            part of the system, it induces changes
            in the density which in turn
            influences the nuclear motion.
            Hence, the electrons mediate between
            the bias and the vibration. The
            feasibility of controlling the nuclear motion
            in a quantum-classical system
            has already been demonstrated. \cite{Castro2014}

            The vibrational degree of freedom is 
            described within the Ehrenfest approximation
            following Verdozzi \textit{et al.} \cite{Verdozzi2006}.
            The modified central part of the electronic
            Hamiltonian reads
            \begin{equation}
                \op{H}_{\textnormal{QD}}(t) = (\epsilon_{\textnormal{QD}} + \lambda x(t)) \sum_{\sigma\in\{\up, \down \}} \op{d}_{\sigma}^\dagger\op{d}_{\sigma}.
            \end{equation}
            The parameter $\lambda$ determines the 
            interaction strength between the 
            electronic and the nuclear system.
            The equation of motion for the vibrational coordinate $x(t)$ is
            \begin{align}
                m\partial_t^2 x(t) &= -\frac{\,d}{\,dx} \left( \frac{1}{2}m\omega^2 x^2 + \langle \Psi | \op{H}_{\textnormal{QD}}(t) | \Psi \rangle \right)\\
                                   &= -m \omega^2 x(t) - \lambda n_{\textnormal{QD}}(t), \\
                x(0)               &= x^0. \nonumber
            \end{align}
            The initial value $x^0$ is calculated self-consistently and the 
            classical equation of motion 
            for the vibrational degree of freedom
            is solved simultaneously with the time-dependent Schr\"odinger equation. 
            The optimization problem for controlling the vibrational coordinate $x(t)$ 
            then reads
            \begin{align}
                \label{eqn:OCP-with-vibrations}
                \min_{\vec{u}_\textnormal{L}, \vec{u}_\textnormal{R} \in \mathbb{R}^{N+1}}   \| x(t) - x^{(\textnormal{target})}(t) \|_{2, [0,T]}^2 \qquad\qquad \quad \ \\
                \nonumber
                \begin{array}{rcl}
                    \textnormal{s.t. }                 & & \\
                    i\partial_t \psi_q(t)              &=& \mathbf{H} (\vec{u}_\textnormal{L}, \vec{u}_\textnormal{R}, x(t), t)\psi_q(t), t \in [0, T],\\ 
                    m\partial_t^2 x(t)                 &=& -m\omega^2 x(t) - \lambda n_{\textnormal{QD}}(t), t \in [0, T],\\
                    \psi_q(0)                          &=& \psi_q^0,\\    
                    x(0)                               &=& x^0,\\
                    \vec{u}_\textnormal{L}             &=& -\vec{u}_\textnormal{R}, \\
                    U_\alpha(\vec{u}_\alpha, \tau_0)   &=& 0,\quad  \alpha \in \{\textnormal{L}, \textnormal{R}\}.
                \end{array}
            \end{align}

\begin{figure}[htb]
                \begin{center}
                    \includegraphics{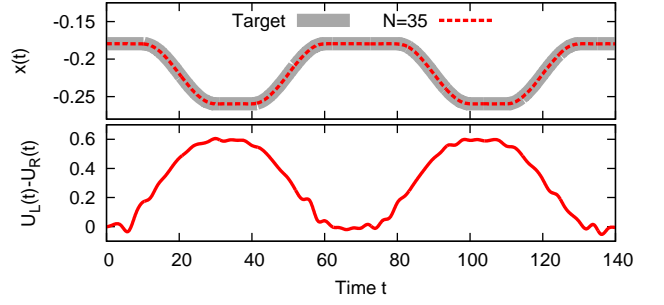}
                    \caption{NQDN junction with an optimized position $x(t)$ of a vibration 
                             coupled to the quantum dot. 
                             The parameters are: $\Gamma_\alpha=0.2, \epsilon_{\textnormal{QD}}=0.5, \xi_\alpha=0, \lambda=0.1, \omega=0.5,m=1$.}
                    \label{fig:NQDN-control-vibration}
                \end{center}
            \end{figure}

            Figure \ref{fig:NQDN-control-vibration} shows the results of such a calculation.

        \subsection{Imposing further constraints on the bias}
        \label{sec3:ssec3}
            In real-world control experiments, an arbitrary
            time-dependence of $U_\alpha(t)$ is difficult
            to achieve. In this section,
            we therefore impose further 
            constraints 
            to restrict the bias $U_\alpha(t)$
            or the derivative $\partial_t U_\alpha(t)$. The optimization problem 
            including such additional constraints then reads
            \begin{align}
                \min_{\vec{u}_\textnormal{L}, \vec{u}_\textnormal{R} \in \mathbb{R}^{N+1}} \| O[\Psi](t) - O^{(\textnormal{target})}(t) \|_{2, [0,T]}^2 \qquad \qquad\\
                \nonumber
                    \begin{aligned}
                        \textnormal{s.t. } \quad i\partial_t \psi_q(t) &= \mathbf{H} (\vec{u}_\textnormal{L}, \vec{u}_\textnormal{R}, t)\psi_q(t), \ t \in [0,T],\\
                        \psi_q(0)                                      &= \psi_q^0,\\
                        U_\alpha(\vec{u}_\alpha, \tau_0)               &= 0,\\
                        \vec{u}_\textnormal{L}                         &= -\vec{u}_\textnormal{R},\\           
                        U_\alpha^{(\textnormal{min})}                               &\le \ \ \ \ U_\alpha(\vec{u}_\alpha, t)           \le U_\alpha^{(\textnormal{max})},\\
                         \widetilde{U}_{\alpha}^{(\textnormal{min})}   &\le \frac{\,d}{\,dt} U_\alpha(\vec{u}_\alpha, t)  \le \widetilde{U}_{\alpha}^{(\textnormal{max})}.
                   \end{aligned}
            \end{align}

            The conditions $U_\alpha^{(\textnormal{min})} \le U_\alpha(\vec{u}_\alpha, t) \le U_\alpha^{(\textnormal{max})}$
            are in general not equivalent to $U_\alpha^{(\textnormal{min})} \le \vec{u}_\alpha \le U_\alpha^{(\textnormal{max})}$,
            unless one uses a monotonic cubic spline. This can be seen in 
            Fig \ref{fig:Spline}, where
            the maximum value of the spline lies between $\tau_1$ and $\tau_2$.
            The constraint for the time derivative is 
            not accessible in this way.
            
\begin{figure}[htb]
                \begin{center}
                    \includegraphics{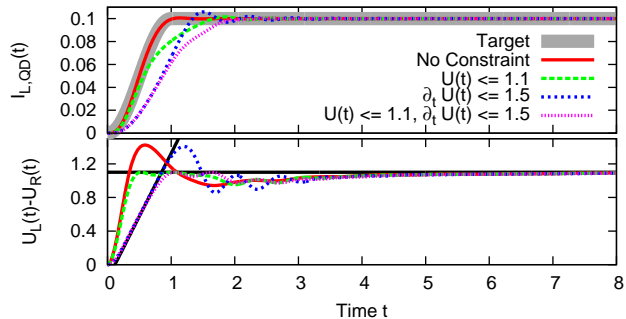}
                    \caption{NQDN junction with an optimized current $I_{\textnormal{L,QD}}(t)$. The 
                             black lines represent the additional constraints $U(t)\le 1.1$
                             and $\partial_t U(t) \le 1.5$.
                             The parameters are: $\Gamma_\alpha=0.2, \epsilon_{\textnormal{QD}}=0.5, \xi_\alpha=0, N=25$.}
                    \label{fig:NQDN-control-current-with-constraints}
                \end{center}
            \end{figure}

            The cubic spline is a third degree polynomial between two nodes $\tau_j$ and
            $\tau_{j+1}$. 
            Thus, the minimum and maximum values can be calculated analytically 
            in every interval $[\tau_j,\tau_{j+1}]$. 
            The constraints are replaced by
            \begin{align}
                \max_{t \in [\tau_j, \tau_{j+1}]} U_\alpha(\vec{u}_\alpha, t) & \le U_\alpha^{(\textnormal{max})},\\
                \min_{t \in [\tau_j, \tau_{j+1}]} U_\alpha(\vec{u}_\alpha, t) & \ge U_\alpha^{(\textnormal{min})},\\
                \max_{t \in [\tau_j, \tau_{j+1}]} \frac{\,d}{\,dt}U_\alpha(\vec{u}_\alpha, t) & \le \widetilde{U}_{\alpha}^{(\textnormal{max})},\\
                \min_{t \in [\tau_j, \tau_{j+1}]} \frac{\,d}{\,dt}U_\alpha(\vec{u}_\alpha, t) & \ge \widetilde{U}_{\alpha}^{(\textnormal{min})}
            \end{align}
            for $j \in \{0, \ldots N-1\}$. Figure \ref{fig:NQDN-control-current-with-constraints} shows the influence of 
            the additional constraints. 
            They are chosen such that the steady state value can still be reached,
            but the transient time is lengthened.

        \subsection{Generating DC currents in Josephson junctions}
            When making the leads superconducting, a junction with an applied DC bias
            does not reach a steady state anymore, but 
            ends up in a time-periodic state. A DC current,
            on the other hand,
            can flow through the junction
            even without applying a bias.
            These phenomena are known as the 
            AC and DC Josephson effects \cite{Josephson1962}.
            The underlying relation is
            \begin{align}
                U(t)    &= \frac{\hbar}{2e} \frac{\,d\chi}{\,dt},\label{eqn:Bias-phase-relation}\\
                \chi(0) &= \chi_{\textnormal{R}} -  \chi_{\textnormal{L}},\\
                I(t)    &= I_0 + I_1\sin(\chi(t))+ I_2\cos(\chi(t)), \label{eqn:Current-phase-relation}
            \end{align}
            where the variables $\chi_\alpha$ describe the 
            phase of the superconducting 
            wave function in lead $\alpha$.
            Thus, the current oscillates with the frequency $\omega=\frac{2e}{\hbar}U$
            when applying a constant bias $U$ across the junction. 
            The values of $I_0, I_1$ and $I_2$
            depend on the bias and only $I_1$ is non-zero for zero bias.
            Following these equations, the only solution for a 
            DC current flowing through the junction would be $\chi(t) \equiv \textnormal{const}$
            and hence $U(t)=0$. But these equations do not take
            switching effects into account and only approximate the current after the 
            transients. 
            In order to force the current to follow a predefined pattern,
            one can make use of the reaction of the current
            to time-dependent changes in the bias. These can be 
            used, for example, to
            compensate the Josephson oscillations.

\begin{figure}[htb]
                \begin{center}
                    \includegraphics{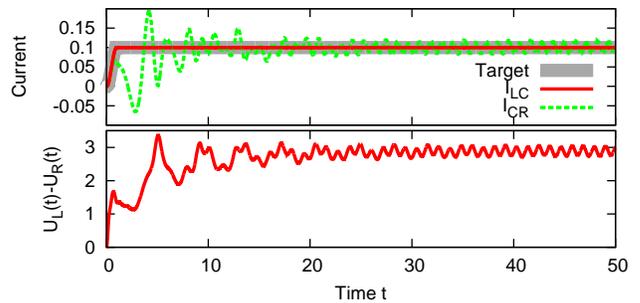}
                    \caption{SQDS junction with an optimized current for two 
                                different number of spline nodes $N$. 
                                The parameters are: $\Gamma_\alpha=0.2, \epsilon_{\textnormal{QD}}=0.5, \xi_\alpha=1, \chi_\alpha=0$.}
                    \label{fig:SQDS-control-current-ILC}
                \end{center}
            \end{figure}

            We start again with optimizing the current $I_{\textnormal{L}, \textnormal{\textnormal{QD}}}(t)$ from the left lead onto 
            the quantum dot such that it follows the target pattern. 
            In this way, we generate a DC current $I_{\textnormal{L}, \textnormal{\textnormal{QD}}}(t)$.
            But the current $I_{\textnormal{QD},\textnormal{R}}(t)$ still shows
            the typical oscillation as it is shown in Fig \ref{fig:SQDS-control-current-ILC}.
        
            In order to obtain a real DC current flowing through
            the Josephson junction, one has to modify the objective function. 
            The idea is to optimize $I_{\textnormal{L}, \textnormal{\textnormal{QD}}}(t)$ and $I_{\textnormal{QD},\textnormal{R}}(t)$ simultaneously such that
            each of them follows a target pattern. 
            The targets have to be chosen carefully, since one might
            end up in situations where the targets cannot be reached
            simultaneously.

            Suppose that the currents $I_{\textnormal{L}, \textnormal{\textnormal{QD}}}(t)$ and $I_{\textnormal{QD},\textnormal{R}}(t)$ follow the 
            predefined patterns perfectly. The density on the quantum dot can then be obtained by integrating 
            the continuity equation at the quantum dot:
            \begin{equation}
                n_{\textnormal{QD}}(t) = n_{\textnormal{QD}}(0) + \int_0^t \,dt' \sum_{\alpha \in \{\textnormal{L},\textnormal{R}\}} I_{\alpha, \textnormal{QD}}^{(\textnormal{target})}(t').
            \end{equation}
            As we see, this can easily lead to contradictions like $n_{\textnormal{QD}}(t)<0$ or $n_{\textnormal{QD}}(t)>2$,
            if the targets are not chosen carefully.
            Even situations with $I_{\textnormal{L}, \textnormal{\textnormal{QD}}}(t) = -I_{\textnormal{R},\textnormal{QD}}(t) \neq 0$ for all times $t$ are in general not possible,
            since the density in such cases would be constant, but switching on a bias 
            normally changes the density. 

\begin{figure}[htb]
                \begin{center}
                    \includegraphics{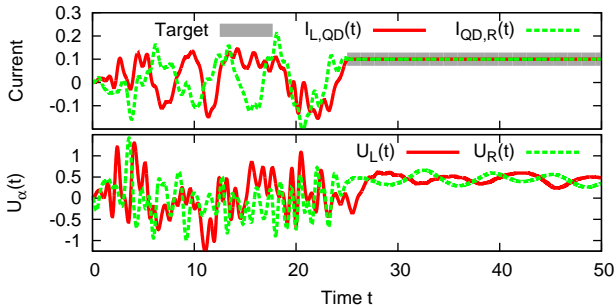}
                    \caption{SQDS junction with optimized currents $I_{\textnormal{L},\textnormal{QD}}(t)$ and $I_{\textnormal{QD},\textnormal{R}}(t)$. We remove 
                            the constraint $U_\textnormal{L}(t)=-U_\textnormal{R}(t)$
                            since the target 
                            can not be reached otherwise. The target 
                            is the same for both currents and starts at $t=25$.
                            The solution exploits the DC Josephson effect.
                            The parameters are: $\Gamma_\alpha=0.2, \epsilon_{\textnormal{QD}}=0.5, \xi_\alpha=1, \chi_\alpha=0, t_0=25, t_1=50$.}
                    \label{fig:SQDS-control-current-both-v1}
                \end{center}
            \end{figure}

\begin{figure}[htb]
                \begin{center}
                    \includegraphics{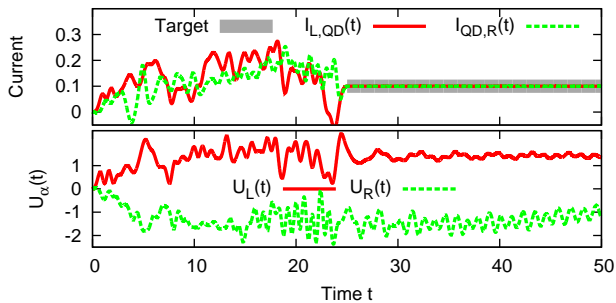}
                    \caption{Same junction as in Fig \ref{fig:SQDS-control-current-both-v1}, 
                             but a different solution to the problem.
                             This solution exploits the AC Josephson effect.}
                    \label{fig:SQDS-control-current-both-v2}
                \end{center}
            \end{figure}

            We avoid these difficulties by using the norm $L^2([t_0,t_1])$, 
            $0\le t_0 <t_1\le T$ in the objective function, which
            is denoted by $\| \cdot \|_{2,[t_0,t_1]}$.
            Furthermore, we remove the constraint $U_{\textnormal{L}}(t)=-U_{\textnormal{R}}(t)$ in order
            to make the targets reachable. 
            The modified optimization problem reads
            \begin{align}
                \label{eqn:OCP-with-constraints}
                \min_{\vec{u}_\textnormal{L}, \vec{u}_\textnormal{R} \in \mathbb{R}^{N+1}}  & \left(\| I_{\textnormal{L},\textnormal{QD}}[\Psi](t) - I_{\textnormal{L},\textnormal{QD}}^{(\textnormal{target})}(t) \|_{2,[t_0,t_1]}^2 \right.\\
                \nonumber                                         & \left. + \| I_{\textnormal{QD},\textnormal{R}}[\Psi](t) - I_{\textnormal{QD},\textnormal{R}}^{(\textnormal{target})}(t) \|_{2,[t_0,t_1]}^2 \right)\\
                \nonumber
                \begin{matrix}
                    \textnormal{s.t.}\\
                    \ \\
                    \
                \end{matrix}
                \quad &  
                \begin{array}{rcl}
                    i\partial_t \psi_q(t)            &=& \mathbf{H} (\vec{u}_\textnormal{L}, \vec{u}_\textnormal{R}, t)\psi_q(t), \ t \in [0,T],\\
                    \psi_q(0)                        &=& \psi_q^0,\\        
                    U_\alpha(\vec{u}_\alpha, \tau_0) &=& 0,\quad  \alpha \in \{\textnormal{L}, \textnormal{R}\}.
                \end{array}
            \end{align}
            The system has now the freedom to adjust the density and currents
            from time $0$ to $t_0$ such that the target patterns can be reached.
            There are two ways to achieve a DC current flowing through a Josephson junction:
            \begin{enumerate}
                \item Following the equations (\ref{eqn:Bias-phase-relation}) - (\ref{eqn:Current-phase-relation}),
                      only the case
                      $U(t)=0$ produces a DC current, namely $I(t) = I_1\sin(\chi_0)$.
                      This is the DC Josephson effect. In general,
                      this relation is not true for our model, since the quantum dot
                      always supports two Andreev bound states for $U=0$ \cite{Stefanucci2010}. They lead to 
                      persistent oscillations in the current and density \cite{Stefanucci2007b, Khosravi2008, Stefanucci2010}.
                      The oscillations in the current can be compensated by small variations of 
                      the bias $U(t) = U_{\textnormal{L}}(t)-U_{\textnormal{R}}(t)$ around the origin. Figure \ref{fig:SQDS-control-current-both-v1}
                      shows an example of such a solution. This approach
                      is limited by $I_1$ and hence does not work for 
                      arbitrary large DC currents.

                \item An alternative approach is to apply a DC bias across the junction, leading to
                      a linear increase in the phase difference $\chi(t)$ and thus to
                      oscillations in the currents. This is the AC Josephson effect.
                      These oscillations can be compensated again by small variations
                      in the bias, the reaction to these changes cancels the Josephson oscillations.
                      Figure \ref{fig:SQDS-control-current-both-v2} shows an example for this type of solutions.
            \end{enumerate}

    \section{Optimizing the Cooper pair splitting efficiency}
    \label{sec:optimize-splitting}

        In this section, we demonstrate how to
        optimize the Cooper pair splitting
        efficiency in a two-quantum dot 
        Y-junction. The overall idea
        is to create entangled electrons
        at two quantum dots.
        
        The entanglement of
        quantum particles has 
        fascinated the scientific community
        since the proposition of the Einstein-Podolsky-Rosen (EPR)
        Gedankenexperiment \cite{Einstein1935}. 
        Entanglement means that two
        particles are linked such that the measurement
        of one particle is sufficient to completely
        determine the quantum state of the other one.
        A prominent example is a pair of electrons 
        with opposite spin.
        Suppose, you have a pair of 
        entangled electron in a spin singlet.
        Then, one spin is up and the other spin is 
        always pointing downwards.
        Photons are a second example
        which can be entangled with respect to 
        the polarization.

        The EPR Gedankenexperiment
        is directly linked to
        the question of
        non-locality of 
        quantum mechanics:
        Can a measurement at 
        position $x$ have an influence
        on a simultaneous or later independent measurement at a 
        different position $x'$? 
        This question can be cast into a
        mathematical formula known
        as Bell's inequality \cite{Bell1964}. 
        A violation of the latter
        would prove the non-locality of 
        quantum mechanics.

        Great progress has been 
        achieved with entangled photons,
        but the final experiment 
        ruling out all possible loopholes
        has not yet been accomplished \cite{Giustina2013}.
        For example, the two measurements
        at $(x,t)$ and $(x',t')$ have 
        to be separated such that
        $c|t-t'| < \|x-x'\|$, i.e. no information
        of the first measurement can be 
        transmitted to the second. Hence 
        large distances are typically required
        to close this loophole \cite{Weihs1998}.
        Another important loophole stems from
        the detector efficiency, i.e. one has to 
        take into account that undetected particles
        might behave completely different compared
        to the detected ones. Typically, one uses the
        fair sampling assumption stating that
        the detected particles are selected randomly 
        and behave statistically in the same way as the undetected ones.

        To do similar Bell test experiments
        with electrons
        is much more difficult and 
        remains an open challenge.
        In recent years,
        a number of ingenious 
        experiments to create
        entangled
        electrons have been performed
        \cite{Hofstetter2009,Hofstetter2011,Herrmann2010,Schindele2012},
        going along with several 
        theoretical developments \cite{Recher2001,Recher2002,Sauret2004,Morten2006,Golubev2007,Burset2011}.
        The basic idea is to use 
        a superconductor as a
        source of entangled electrons.
        In the BCS ground state, 
        electrons form Cooper pairs
        due to the attractive interaction
        caused by phonons. These pairs
        consist of two electrons 
        with opposite spin and momentum.
        
\begin{figure}[htb]
            \begin{center}
                \includegraphics{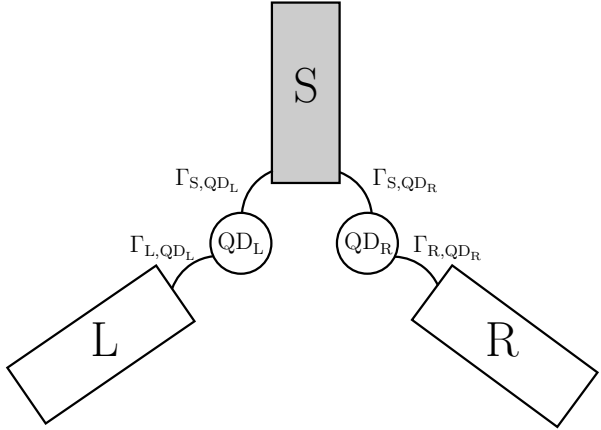}
                \caption{Sketch of the Y junction and explanation of 
                        all relevant parameters. Only
                        the lead labeled with S is superconducting.
                        The gray color is used to indicate
                        the superconducting part.
                        The aim is to create entangled
                        electrons on the two quantum dots.
                        All three leads are 
                        semi infinite.
                        }
                \label{fig:Sketch-Y-junction}
            \end{center}
        \end{figure}

        The idea is to create a splitted
        Cooper pair at the two 
        quantum dots, i.e. one 
        electron is on the left
        quantum dot and the other with opposite spin is on
        the right one (see sketch in Fig. \ref{fig:Sketch-Y-junction}).
        However, this process
        competes with the case
        of both electrons moving onto the same quantum dot.
        The latter can be suppressed 
        by a large charging energy 
        of the quantum dots caused by
        the Coulomb interaction. 
        This make double occupancies less likely.
        
        We propose a way 
        to achieve splitting 
        efficiencies of $99\%$ and more,
        which we hope will 
        help the eventual experimental
        demonstration of the violation
        of Bell's
        inequality. 
        In comparison to traditional
        approaches, our method has two major 
        differences. First, 
        we do not rely on a 
        large Coulomb repulsion
        on the quantum dots
        but rather use optimal control theory
        to tailor the bias in the normal 
        leads in such a way that the splitting
        probability is maximized.
        Second, we look at the
        Cooper pair density on the 
        quantum dots as opposed to
        the experimental approaches working 
        currents of entangled electrons in the two
        normal conducting leads.
        Consequently, a direct 
        comparison of results is not
        easily possible as the 
        efficiencies measure different
        ratios.
        As a future work, it might be worth doing
        an extensive comparative study
        answering whether the here created pair
        eventually moves towards the leads 
        or stays on the quantum dots.
        In experiments, splitting efficiencies
        for the current of $90\%$ have been 
        realized in recent experiments \cite{Schindele2012}
        being significantly higher than previous results.
        Despite this progress,
        the experimental proof 
        of the violation of
        Bell's inequality 
        is still pending.

        In contrast to all 
        systems studied in the previous sections, we now work 
        with three leads.
        The system is sketched in Fig \ref{fig:Sketch-Y-junction}.
        It consists of two quantum dots 
        ($\textnormal{QD}_\textnormal{L}$ and $\textnormal{QD}_\textnormal{R}$),
        one superconducting ($\textnormal{S}$) and two normal
        leads ($\textnormal{L}$ and $\textnormal{R}$). 

        The Hamiltonian of our modified model reads
        \begin{align}
            \label{eqn:sec4:full-Hamiltonian-Y-junction}
            \op{H}(t)            &= \sum_{\alpha \in \{\textnormal{L},\textnormal{R},\textnormal{S}\}} \op{H}_{\alpha} + \sum_{\alpha \in \{\textnormal{L},\textnormal{R},\textnormal{S}\}}\op{H}_{T,\alpha}(t),\\
            \op{H}_{\alpha}              &= \sum_{k=0}^\infty \sum_{\sigma\in\{\up,\down\}}\left(t_\alpha \op{c}_{\alpha k\sigma}^\dagger \op{c}_{\alpha(k+1)\sigma} + H.c. \right)\\
                                         &  \nonumber\qquad + \sum_{k=0}^\infty \left(\Delta_\alpha e^{i\chi_\alpha} \op{c}_{\alpha k\up}^\dagger \op{c}_{\alpha k\down}^\dagger + H.c. \right),
        \end{align}    
        \begin{align}
            \op{H}_{T,S}(t)              &= \sum_{\alpha \in \{\textnormal{L},\textnormal{R}\}}\sum_{\sigma \in \{\up,\down\}}\left( t_{\textnormal{S}, \textnormal{QD}_\alpha} \op{c}_{\textnormal{S}0\sigma}^\dagger \op{d}_{\textnormal{QD}_\alpha \sigma} + H.c. \right),\\
            \op{H}_{T,\alpha}(t)         &= \sum_{\sigma \in \{\up,\down\}}\left( t_{\alpha, \textnormal{QD}_\alpha} e^{i\gamma_{\alpha, \textnormal{QD}_\alpha}(t)}\op{c}_{\alpha0\sigma}^\dagger \op{d}_{\textnormal{QD}_\alpha \sigma} + H.c. \right) \nonumber \\
                                         & \qquad \qquad \qquad \textnormal{for }\alpha \in \{ \textnormal{L}, \textnormal{R} \}.
            \label{eqn:sec4:last-eqn-full-Hamiltonian-Y-junction}
        \end{align}

        Note that there is only a bias in the left and right lead.
        All parameters are again chosen real and positive.
        Furthermore, we work at temperature $T=0$ and assume the wide band limit $t_{\alpha, \textnormal{QD}_\beta} \ll t_\alpha $.     
        Again, only the coupling strengths
        $\Gamma_{\alpha, \textnormal{QD}_\beta} = {2t_{\alpha, \textnormal{QD}_\beta}^2}/{t_\alpha}$ will be stated.
        
\begin{figure}[htb]
            \begin{center}
                \includegraphics{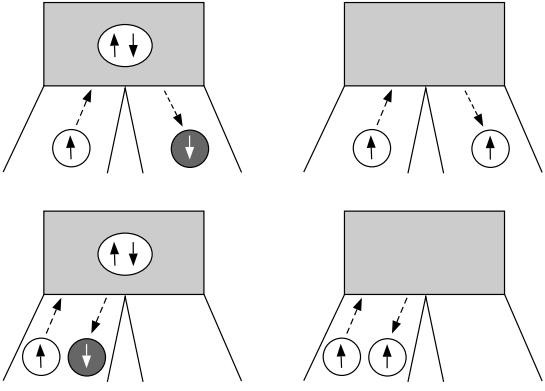}
                \caption{Overview of the four possible reflection processes.
                         Black arrows indicate electrons, white arrows
                         represent holes. The gray block is the superconducting lead 
                         $\textnormal{S}$ of Fig. \ref{fig:Sketch-Y-junction}.
                         Top left: Sketch of a crossed Andreev reflection.
                         The incoming spin up electron in the left lead gets 
                         reflected as a spin down hole to the right lead.
                         Simultaneously, a Cooper pair is created in the 
                         superconducting lead. The opposite process, which
                         removes a Cooper pair from the superconductor,
                         is also possible.
                         Bottom left:
                         The reflected hole stays in the left lead.
                         This corresponds to the normal Andreev reflection.
                         Top right: Sketch of an elastic cotunneling process.
                         Now, the incoming electron gets reflected into 
                         the right lead. 
                         Bottom right: Alternatively, the 
                         electron can also be reflected into the 
                         left lead corresponding to normal reflection.}
                \label{fig:Crossed-Andreev-reflection}
            \end{center}
        \end{figure}

        In the following, we demonstrate how to
        optimize the Cooper pair splitting
        efficiency in the above model
        of a two-quantum dot Y-junction. 
        The goal is to 
        operate the device as
        a Cooper pair splitter
        that creates
        entangled electrons
        on the two quantum dots.
        The splitting of a Cooper pair 
        can be understood
        as a crossed Andreev reflection.
        An incoming electron in one of the
        normal leads gets reflected
        into the other lead as a hole.
        This creates a Cooper pair 
        in the superconductor. The
        process is sketched in Fig. \ref{fig:Crossed-Andreev-reflection} (top left).
        Similarly, the opposite process 
        removes a Cooper pair from the
        superconductor. 
        Besides, there 
        are three other possible
        reflection processes: (a) normal reflection,
        (b) Andreev reflection,
        and (c)
        elastic cotunneling.
        The latter corresponds to
        a reflection of the incoming electron
        to the opposite lead.
        These three processes together with 
        the crossed Andreev reflection 
        are all sketched in
        Fig. \ref{fig:Crossed-Andreev-reflection}. 
        
        The central ingredient for the 
        optimization
        process is the proper definition
        of a suitable 
        objective function
        which is then to be maximized.
        It has 
        to quantify the Cooper pair splitting
        efficiency. 
        To this end, we first define
        the so-called
        pairing density or anomalous density as
        \begin{align}
            P_{\textnormal{QD}_\alpha,\textnormal{QD}_\beta}(t) &= \langle \hat{d}_{\textnormal{QD}_\alpha\downarrow}(t)\hat{d}_{\textnormal{QD}_\beta\uparrow}(t)\rangle.
        \end{align}
        We use its absolute value
        squared $|P_{\textnormal{QD}_\alpha,\textnormal{QD}_\beta}(t)|^2$ 
        as a measure 
        for the Cooper pair density
        with one electron at $\textnormal{QD}_\alpha$
        and the other at $\textnormal{QD}_\beta$.
        We propose to maximize 
        the following 
        objective function:
            \begin{equation}
                \label{eqn:objective-function}
                \frac{1}{t_1-t_0}\int_{t_0}^{t_1} \,dt \frac{\sum_{\alpha \neq \alpha'\in \{\textnormal{L}, \textnormal{R}\}}|P_{\textnormal{QD}_{\alpha},\textnormal{QD}_{\alpha'}}(t)|^2}
                    {\sum_{\alpha,\alpha' \in \{\textnormal{L}, \textnormal{R}\}}|P_{\textnormal{QD}_{\alpha},\textnormal{QD}_{\alpha'}}(t)|^2}.
            \end{equation}
        The fraction represents the
        Cooper pair splitting efficiency at time t, which is expressed 
        as the amount of Cooper pairs being split up divided 
        by the total amount of Cooper pairs on the quantum dots. 
        We calculate its average over the time span 
        from $t_0$ to $t_1$.
        The pairing densities $P_{\textnormal{QD}_\alpha,\textnormal{QD}_\beta}(t)$
        are obtained from the single particle wave functions $\psi_q(t)$, i.e.,
        the solutions of the time-dependent 
        Bogoliubov-de Gennes equation (\ref{eqn:BdG-equation}). 

        We want to tailor the bias such that
        we maximize the time averaged 
        Cooper pair splitting
        efficiency.
        The corresponding optimization problem then reads
            \begin{align}
                &\max_{\vec{u}_{\textnormal{L}}, \vec{u}_{\textnormal{R}} \in \mathbb{R}^{N+1}} 
                    \frac{1}{t_1-t_0}\int_{t_0}^{t_1}\,dt \frac{\sum_{\alpha \neq \alpha'\in \{\textnormal{L}, \textnormal{R}\}}|P_{\textnormal{QD}_{\alpha},\textnormal{QD}_{\alpha'}}(t)|^2}
                    {\sum_{\alpha,\alpha' \in \{\textnormal{L}, \textnormal{R}\}}|P_{\textnormal{QD}_{\alpha},\textnormal{QD}_{\alpha'}}(t)|^2} \nonumber \\
                &\begin{array}{rcl}
                    \textnormal{s.t. } P_{\textnormal{QD}_\alpha, \textnormal{QD}_\beta}(t) &=& \int \,dq f(\epsilon_q) u_q(\textnormal{QD}_\alpha, t) v_q(\textnormal{QD}_\beta, t)^\star,\\
                    \quad i\partial_t \psi_q(t)   &=& \mathbf{H} (\vec{u}_{\textnormal{L}}, \vec{u}_{\textnormal{R}}, t)\psi_q(t), \quad t \in [0, T],\\ 
                    \psi_q(0)                                        &=& \psi_q^0,\\     
                    U_\alpha(\vec{u}_\alpha, \tau_0)                 &=& 0,\quad  \alpha \in \{\textnormal{L}, \textnormal{R}\}.
                \end{array} 
            \end{align}

        The problem can be solved using again standard 
        derivative-free
        algorithms for non-linear optimization problems,
        for example the ones provided by the library NLopt \cite{NLopt_paper}.

        \begin{figure}[htb]
            \begin{center}
                \includegraphics[]{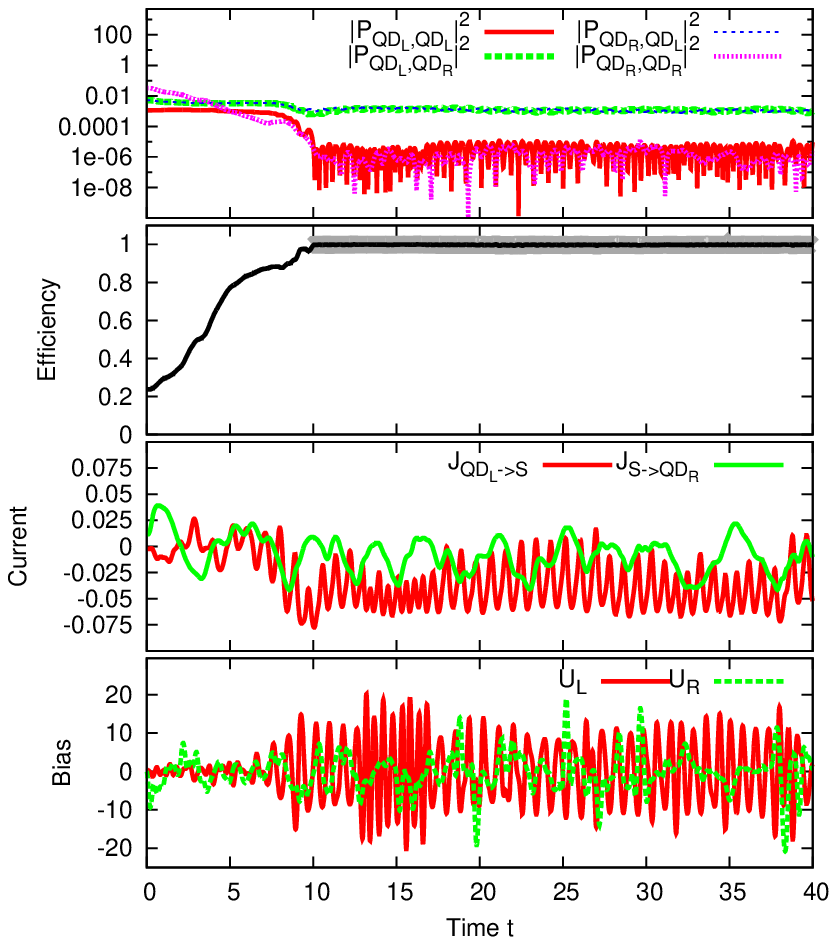}
                \caption{Simulation with an optimized bias. 
                         (a) Top: $|P_{\textnormal{QD}_\alpha, \textnormal{QD}_\beta}(t)|^2$
                         as a function of time. 
                         (b) Second from top: Resulting efficiency,
                         gray line indicates time interval of 
                         optimization.
                         second from bottom
                         (c) Second from bottom: 
                         Resulting currents $I_{\textnormal{QD}_{\textnormal{L}},\textnormal{S}}(t)$
                         and $I_{\textnormal{S}, \textnormal{QD}_{\textnormal{R}}}(t)$. 
                         (d) Bottom: Tailored bias $U_\textnormal{L}(t)$
                         and $U_\textnormal{R}(t)$ of the optimization.
                         The parameters are: 
                         $\Gamma_{\textnormal{S},\textnormal{QD}_\textnormal{L}} = \Gamma_{\textnormal{S},\textnormal{QD}_\textnormal{R}} = \Gamma_{\textnormal{N},\textnormal{QD}_\textnormal{L}}=0.2$,
                         $\Gamma_{\textnormal{N},\textnormal{QD}_\textnormal{R}} = 1$, $\xi_\textnormal{S}=1$, $\xi_\textnormal{L} = \xi_\textnormal{R}=0$, $N=200$.}
                \label{fig:efficiency-optimized}
            \end{center}
        \end{figure}

        To achieve high splitting efficiencies it 
        is essential that the
        junction is asymmetric,
        i.e. the couplings to the left 
        and to the right quantum dot must not be equal.
        This is necessary since we observe
        an upper bound of $50\%$ for the Cooper
        pair splitting efficiency
        in symmetric junctions,
        which is already achieved in 
        the ground state
        by the usual Cooper pair 
        tunneling leading to the proximity effect. Hence any
        optimization starting in the ground
        state will not improve
        the results. The underlying cause
        for this limitation is still unknown
        and under investigation.
        In order to bypass this issue, we
        choose an asymmetric coupling
        of the quantum dots to the normal leads.
        
        The results of such an optimization are depicted in
        Fig. \ref{fig:efficiency-optimized}.
        The bias is tailored such that the Cooper pair 
        splitting efficiency is maximized. It suppresses the non-splitting 
        processes. The efficiency
        is optimized in the time interval
        from $t_0=10$ to $t_1=40$.
        This interval is indicated by the underlying
        thick gray line in the
        plot of the efficiency (second from top).
        In this interval, we 
        achieve an average efficiency of 
        more than $99\%$.
        The values of $|P_{\textnormal{QD}_\textnormal{L},\textnormal{QD}_\textnormal{R}}(t)|^2$
        and  $|P_{\textnormal{QD}_\textnormal{R},\textnormal{QD}_\textnormal{L}}(t)|^2$
        are on top of each other.
        The resulting currents flowing through the 
        junction indicate, that in the time average,
        there is a net current flowing from the
        right normal conducting lead (R) via the
        superconductor (S) to the left one (L).
        This is deduced from the observation
        that $I_{\textnormal{QD}_{\textnormal{L}},\textnormal{S}}(t)$
        and $I_{\textnormal{S}, \textnormal{QD}_{\textnormal{R}}}(t)$
        are both negative in the time average.
        We point out, that this does not
        say anything about the movement of the
        entangled Cooper pairs.
        
        This result clearly demonstrates that the
        Coulomb interaction at the quantum dots
        is not necessary in order to
        obtain high efficiencies. One
        can also succeed with optimized
        biases.

    \section{Conclusion}
        \label{sec:conclusion}
        Usually, in the field of molecular electronics, the goal is to calculate
        the steady-state or time-dependent current that is generated by a
        given bias and gate voltage. Sometimes, however, one may be interested
        in taking a step beyond this point and control the current or other 
        observables of the junction. To this end we have presented an algorithm
        that allows us to calculate the time-dependent bias that achieves a
        prescribed goal. In the examples presented, we determine numerically the 
        time-dependent bias that forces the current, the density or the molecular
        vibration to follow a given temporal pattern. The method is general and
        not restricted to the observables listed above. In the final section we
        apply our approach to optimize the Cooper pair splitting efficiency in
        a Y-junction with two quantum dots. We successfully create spatially separated 
        entangled electron pairs with an efficiency of nearly 100\%.
        We expect our approach to be useful in the control of other - essentially
        arbitrary - observables in molecular junctions.


\bibliography{BibtexDatabase}

\end{document}